\documentclass[journal]{vgtc} 

\usepackage{cite}
\usepackage[bottom]{footmisc}
\usepackage{mathptmx}
\usepackage{graphicx}
\usepackage{siunitx}
\usepackage{times}
\usepackage[T1]{fontenc}
\usepackage{enumitem}
\usepackage[table]{xcolor} 
\usepackage{wrapfig}
\usepackage{multirow}
\usepackage{adjustbox}
\usepackage{mwe}
\usepackage{graphbox}
\usepackage{dirtytalk}
\usepackage{soul}
\usepackage{mdframed}
\usepackage{tabularx}
\usepackage{xcolor}
\usepackage{enumitem}
\usepackage{balance}
\definecolor{CindySalmon}{RGB}{232, 125, 114}
\definecolor{greenB}{RGB}{77, 175, 74}
\definecolor{purpleF}{RGB}{152,78,163}

\newcommand{\pheading}[1]{\smallskip\noindent\textbf{#1}}

\newcommand{\change}[1]{\textcolor{black}{#1}}

\usepackage[bookmarks,backref=true,linkcolor=black]{hyperref} 
\hypersetup{
 pdfauthor = {},
 pdftitle = {},
 pdfsubject = {},
 pdfkeywords = {},
 colorlinks=true,
 linkcolor= black,
 citecolor= black,
 pageanchor=true,
 urlcolor = black,
 plainpages = false,
 linktocpage
}

\onlineid{1151} 
\vgtccategory{Theoretical and Empirical} 
\vgtcinsertpkg 

\title{Same Data, Diverging Perspectives: \\The Power of Visualizations to Elicit Competing Interpretations}

\author{Cindy Xiong Bearfield, Lisanne van Weelden, Adam Waytz, Steven Franconeri}
\authorfooter{ 
\item
Cindy Xiong Bearfield is with the Georgia Institute of Technology.
\item
Lisanne van Weelden is with Utrecht University, the Netherlands.
\item
Adam Waytz is with the Kellogg School of Management, Northwestern University.
\item
Steven Franconeri is with Northwestern University.
}

\shortauthortitle{Xiong \MakeLowercase{\textit{et al.}}: }

\abstract{People routinely rely on data to make decisions, but the process can be riddled with biases. 
We show that patterns in data might be noticed first or more strongly, depending on how the data is visually represented or what the viewer finds salient. 
We also demonstrate that viewer interpretation of data is similar to that of ‘ambiguous figures’ such that two people looking at the same data can come to different decisions. 
In our studies, participants read visualizations depicting competitions between two entities, where one has a historical lead (A) but the other has been gaining momentum (B) and predicted a winner, across two chart types and three annotation approaches.
They either saw the historical lead as salient and predicted that A would win, or saw the increasing momentum as salient and predicted B to win. 
These results suggest that decisions can be influenced by both how data are presented and what patterns people find visually salient.}


\begin{document}
\maketitle

\section{Introduction}
\begin{figure}[th!]
\centering
\includegraphics[width = 0.85\linewidth]{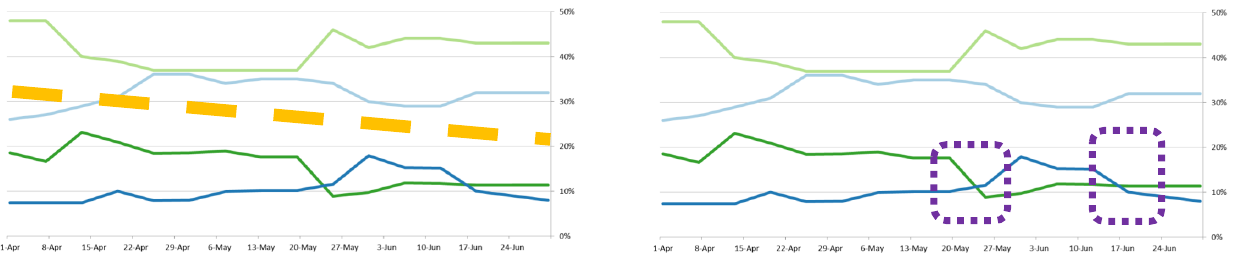}
\caption{Visualizations contain many patterns one could focus on. Visual annotation can guide people's attention to patterns they might otherwise miss. \change{Left: Annotation showing that the green lines are mirror images of each other. Right: Annotation showing that the bottom two lines cross each other twice. Adapted from Xiong et al., see \cite{xiong2019curse}.}}
\label{fig:curse}
\end{figure}

There exist many patterns within a single visualization to perceive.
People can be biased to see different patterns as visually salient. 
For example, in Figure \ref{fig:curse}, readers can find either the symmetry between the two green lines or the intersection between the bottom two lines to be the most visually salient.
Therefore, one principle of effective visualization design is to thoughtfully highlight key patterns you would like a reader to see.
However, each visualization design decision made can guide viewer attention to certain patterns.
Picking a visual representation, such as a chart type, can strongly influence what patterns people see \cite{tversky2014visualizing, cleveland1984graphical, mackinlay1986automating, nothelfer2019measures}. 
It is therefore critical to better understand how visual design affects the \textit{decisions} people make about data. 
We investigate how visualization design choices can alter \textit{predictive decisions} people make with data. 
We compare the effect of several common visual storytelling techniques, from annotation and highlighting to altering the visual representation type, using neutral topics to account for the effect of top-down factors such as people's pre-existing beliefs as much as possible.
The result also sheds light on individual differences in how salient features might drive reader decisions in visual data communication.
\pheading{Contribution:} 
We contribute three empirical experiments demonstrating that visualization design (i.e., visual representation and storytelling techniques) can change perceived patterns in data and people's decision. 
\change{We showed participants bar charts, tables, and line charts and asked them to report salient patterns they see in the data and predict future data trends.}
\change{We tested two storytelling techniques: visual annotation, which involved directly drawing bounding boxes and trend lines on the visualizations, and visual highlighting, which involved manipulating certain colors in the visualization to increase the perceived saliency of the corresponding data values.}
\change{We found evidence supporting visual annotation as more effective than visual highlighting in swaying reader decisions.}
This provides empirical evidence for visualizations as rhetorical devices, supplementing existing work in visualization sciences.
In many scenarios, a visualization might be the only contact that an information consumer has with the underlying data \cite{correll2019ethical}. 
The story the visualization designer chooses to tell can profoundly impact on people's decisions and actions \cite{barocas2017engaging, black2012ibm, holder2022dispersion}.  
Based on the results of this work, \change{we advocate for the visualization research community to reflect upon the ethical implications of data storytelling and consider future steps to generate ethical storytelling guidelines to encourage readers to think critically when reading visualizations.}

\section{Background}



A dataset can contain many perceivable patterns. 
When reading a visualization, viewers need to exercise top-down attentional control to extract a series of alternative relationships and patterns from data values \cite{szafir2016four, egeth2010salience, michal2016visual, michal2017visual}. 
For example, when a scientist reads a bar chart showing two groups of two bars, they could extract main-effect comparisons like ``overall, people performed better in condition A than condition B'' or interactions like ``group X performed better in condition A but group Y performed better in condition B'' \cite{shah2011bar, xiong2021visual, bearfield2023does}. 
We compare this characteristic of visualizations to `ambiguous figures' from cognitive psychology.
While the components of a visualization accurately represent the underlying data, user interpretation of the charts can be ambiguous and multi-stable, similar to interpreting an ambiguous figure like the duck-rabbit or the Necker cube \cite{attneave1971multistability}.

Existing work has demonstrated that the multi-stability of visualization interpretations can lead to different data pattern recognition and decisions due to people's existing beliefs, knowledge, and motivation.
For example, when participants were primed with knowledge of certain patterns in a line chart depicting fictional political voting data, they began to find those patterns more visually salient \cite{xiong2019curse}.
When viewing a visualization depicting global temperature trends, viewers who believed in climate change tended to look at the increasing sections, while other viewers who did not believe in climate change tended to look more at the flat sections, suggesting that they were confirming their preconceptions~\cite{luo2019motivated}.
On social media, people can draw significantly different inferences from similar data depending on whether they hold beliefs related to pro- or anti-masking \cite{lee2021viral}.
We propose that the multi-stability in visualization interpretation can also happen in more controlled settings in which people are reading visualizations not explicitly associated with prior beliefs, which we demonstrate in Experiments 1a and 1b.

We also hypothesize that the bottom-up \textit{visual saliency} of a pattern in a visualization and \textit{individual differences} \change{might be able to influence what patterns people tend to seek out in data.}
Existing work from human perception demonstrates that in visual search tasks, salient targets pop out, while salient distractors capture attention and make search more difficult \cite{theeuwes1991cross, theeuwes2010top}.
Saliency models can predict where people look in a visualization as a form of quantitative visual attention model \cite{bylinskii2017understanding}. 
Visual features such as hot spots, titles, and image centers tend to attract people's gaze \cite{janicke2010salience, matzen2017data, bylinskii2017understanding}. 
Salient colors attract viewer attention and can make certain data points `pop out' \cite{healey2011attention}. 
Yet little work has explored whether salience can actually affect the downstream \textit{decisions} people make in addition to where they look.
We extend this work by examining how increased visual salience in data patterns can influence what patterns people extract from a visualization and what decisions they make (see Experiment 3).

On the individual difference front, researchers studying asset markets have modeled how investors extract trends and form expectations about data patterns.
For example, people seem to hold one of three types of \textit{expectations} when making financial decisions \cite{dominitz2011measuring}: \textbf{persistence} type investors believe that trends will persist into the future, \textbf{mean-reversion} type investors believe that trends will reverse in the near future, and \textbf{random-walk} type investors believe that financial returns do not follow a specific pattern and are independently and identically distributed over time. 
We hypothesize that these three expectation types might generalize to non-finance data interpretation and decision-making. 
People may possess traits that make them pick out similar patterns and make similar decisions even across two visualizations depicting different topics, corroborating the series of visualization studies that observed an effect of individual differences on visual data interpretation (e.g., \cite{schneider2017your, alves2020exploring}).
\change{For example, Ziemkiewicz et al.~\cite{ziemkiewicz2012visualization} demonstrated that} visualization design components such as color, interaction, and labeling can impact interpretation accuracy, reading speed, and preferences depending on viewers' locus of control and personality factors. 
In a study that examined how visualization complexity impacts perceived trust, perceived transparency, and preferences for map-based visualizations, researchers found two types of participants: ones that prefer and trust simple visualizations that prioritize clarity, and ones that prefer and trust complex visualizations that prioritize thoroughness \cite{xiong2019examining}.
These results motivate us to test for potential intra-personal consistencies in pattern extraction and data interpretation (see Experiment 2).
\subsection{Visualization Affordance}
Visual design decisions can impact what people see in data and how they reason with data \cite{xiong2022reasoning, holder2022dispersion}.
Chart type, for example, can change one's intuitive reaction to data.
Line charts make trends and correlations easy to detect, while scatter plots facilitate spotting outliers \cite{zacks1999bars}. 
Using spatial location to encode data values can increase the accuracy of data value perception \cite{heer2010crowdsourcing}, but using area, hue, and color intensity (e.g., heat maps) can provide useful `big picture' information \cite{albers2014ensemble}. 
Histograms help people find extremes, scatterplots make clusters more salient, choropleth maps facilitate comparisons of approximate values, and treemaps encourage identification of hierarchical structures \cite{lee2016vlat, mackinlay1986automating}.
Recent work suggests that these tendencies are driven by perceptual processes.
When people make sense of visualizations, the brain extracts data patterns using perceptual proxies, shortcuts that identify the `gist' from visual marks such as the shape of the line or the center of mass for a scatterplot \cite{jardine2019perceptual, ondov2020revealing}.
These processes create perceptual affordances. 
For example, when designing icon arrays to communicate probabilistic information, the arrangements of the icon grids can bias people to perceptually misestimate the probabilities \cite{xiong2022investigating}. 
People can underestimate average values in line charts and overestimate average values in bar charts \cite{xiong2019biased}, though the amount of overestimation depends on the aspect ratio of the bars \cite{ceja2020truth}.
Depending on how data values are spatially arranged in a bar chart, people compare different sets of values in their takeaways \cite{xiong2021visual, gaba2022comparison}.


Visualization design also profoundly impacts viewers' thinking patterns. 
For example, presenting highly aggregated data elicits causal interpretations while presenting dis-aggregated data elicits more appropriate correlational interpretations \cite{xiong2019illusion}.
Charts that show probabilistic outcomes as discrete objects can promote a better understanding of uncertainties and effect sizes \cite{kay2016ish, hawley2008impact, kale2020visual, hofman2020visualizing}.
We contribute to literature by further showing that design decisions can impact the \textit{decisions} people make in data (see Experiment 1a).

\subsection{Data Storytelling Techniques}

Extracting one key pattern of the many in a visualization can be inefficient \cite{szafir2016four, shah2011bar}, especially when the data is complex. 
Narrative visualizations \cite{segel2010narrative} and data storytelling techniques, such as interactive narration \cite{shi2022breaking, feng2018effects}, animation \cite{shin2022roslingifier}, verbal introductions \cite{yang2021explaining}, and data comics \cite{wang2019comparing}, help designers share data insights in more engaging and memorable ways \cite{lee2015more, hullman2013deeper, stolper2016emerging}.
Practitioners identified the `focusing’ technique to be among the most common and effective \cite{ajani2021declutter}.
In this technique, the key patterns a viewer should extract and remember from a visualization is \textit{annotated} and the values that create this key pattern are \textit{highlighted} \cite{hullman2011visualization, vora2019power, segel2010narrative, jones2020, andrews2019info}.

\textbf{Annotations} have been used widely to help readers more efficiently take away key messages or notice relevant patterns \cite{stokes2022balance, kim2021towards}. 
Although annotating may not align with the minimalist design philosophies of Tufte \cite{tufte2001visual}, recent work suggests that many find it effective depending on the content of the annotation and user goals \cite{stokes2022balance}.
The title is a common and effective annotation used in visualizations.
Titles have been shown to enhance objective memory of visualized content \cite{borkin2015beyond}.
The framing of the title can lead to different perspectives \cite{kong2019trust, kong2018frames}.
In one study, researchers created titles that are slanted, open-ended, or simply describing statistics, and found that slanted titles can drive viewers to extract opposing messages from the same visualization \cite{kong2018frames}.
In a follow-up study, Kong et al.~\cite{kong2019trust} found the biasing effect of slanted titles to persist even when the content of the title and visualization are misaligned.
Participant recall of the visualization's key message was also more aligned with the title than the visualization.
We build on this work by first showing that even without titles nudging people, two people can look at the same visualization and draw different conclusions.
We bridge the gap by looking at the effect of \textit{non-verbal} annotations on influencing the \textit{patterns} readers extract and the \textit{decisions} they make.
This will help us observe the extent to which past findings regarding titles and slants generalize to other storytelling techniques and identify a wider set of reader behaviors that could be impacted.

\textbf{Highlighting} key patterns is another popular technique to help people focus on particular patterns in data, drawing attention to a certain location on the visualization through a bottom-up process (also referred to as stimulus-driven attention, see \cite{chun2011taxonomy, yarbus2013eye}).
Highlighting can guide people to see patterns they might otherwise miss \cite{xiong2019curse} by increasing the visual saliency of key patterns and reducing the `noise' from other competing patterns \cite{hill2018minimalism, inbar2007minimalism, ajani2021declutter, andrews2019info}.
Visualizations that highlight relevant values were rated as more clear and aesthetically pleasing, and key messages were more accurately remembered than in their non-highlighted counterparts \cite{ajani2021declutter}.
We expand upon existing work by comparing the strength of annotation and highlighting on impacting the patterns people see and the decisions they make (see Experiment 3).
We compare the competing strength in bottom-up cues to emphasize patterns in data and top-down tendencies people have to extract certain patterns from data to identify more effective storytelling techniques and derive practical and ethical implications of data storytelling.

\section{Experiment Hypotheses and Overview}
\label{Hypotheses}
We conducted four experiments demonstrating the power of visualization design to elicit different interpretations from visualization readers. 
All statistical analyses were done in R. The data and R-script used for analyses are available at the Open Science Framework website: \url{https://osf.io/vmnh5}. We make several hypotheses based on related work. \change{We present the alternative hypotheses tested below for clarity. The null hypothesis is that we do not observe these effects.}

\pheading{Hypothesis No.1.} Visualizations can be interpreted as `ambiguous figures' such that readers can see different patterns in the same visualization and thus make different decisions. (See Experiment 1)

\pheading{Hypothesis No.2.} Visualization design impacts the salient features people see in data, and subsequently the decisions they make with data. Specifically, bar charts and table designs can change the saliency of data patterns to drive different interpretations. (See Experiment 1)

\pheading{Hypothesis No.3.} There exist some intra-personal consistencies when people read visualizations, such that the patterns and decisions people see in one chart can transfer to influence the patterns they see in a subsequent chart and their decisions. (See Experiment 2). 

\pheading{Hypothesis No.4.} The visual saliency of patterns in a visualization can influence what people see and their decisions (See Experiment 3). 

\pheading{Hypothesis No.5.} Storytelling techniques such as annotation and highlighting/recoloring can bias visualization readers to not only see different patterns but also lead to different decisions. Furthermore, these techniques vary in effectiveness such that annotation and highlighting are not equally effective. (See Experiment 3). 
\section{Experiment 1a: Table and Bar Chart Affordances}

Experiment 1 shows that data visualizations are not agnostic.
Rather, people's interpretation can be multi-stable, similar to `ambiguous figures' from cognitive psychology such that two people looking at the same information can see different patterns and come to different conclusions.
We created a simple, neutral dataset visualized as a bar chart and demonstrated that participants can see different patterns as visually salient and draw different conclusions (Hypothesis 1).
We compare the distribution of participant conclusions when reading a bar chart and reading a table presenting the same information and demonstrate that visualization design (bar versus table) can afford different viewer conclusions, even within a simple dataset (Hypothesis 2).

\subsection{Participants, Design, and Procedure}
We recruited 132 participants from Amazon's Mechanical Turk. 
After filtering for workers who have an approval rate above 95\%, passed our attention checks (e.g., select this option if you are paying attention), and did not enter non-sensical responses, we are left with 125 individuals ($M_{age}$ = 32.84(8.81), 51 female). 
The participants were compensated at the rate of 9 dollars per hour.

\begin{figure}[th!]
\centering
\includegraphics[width = 0.86\columnwidth]{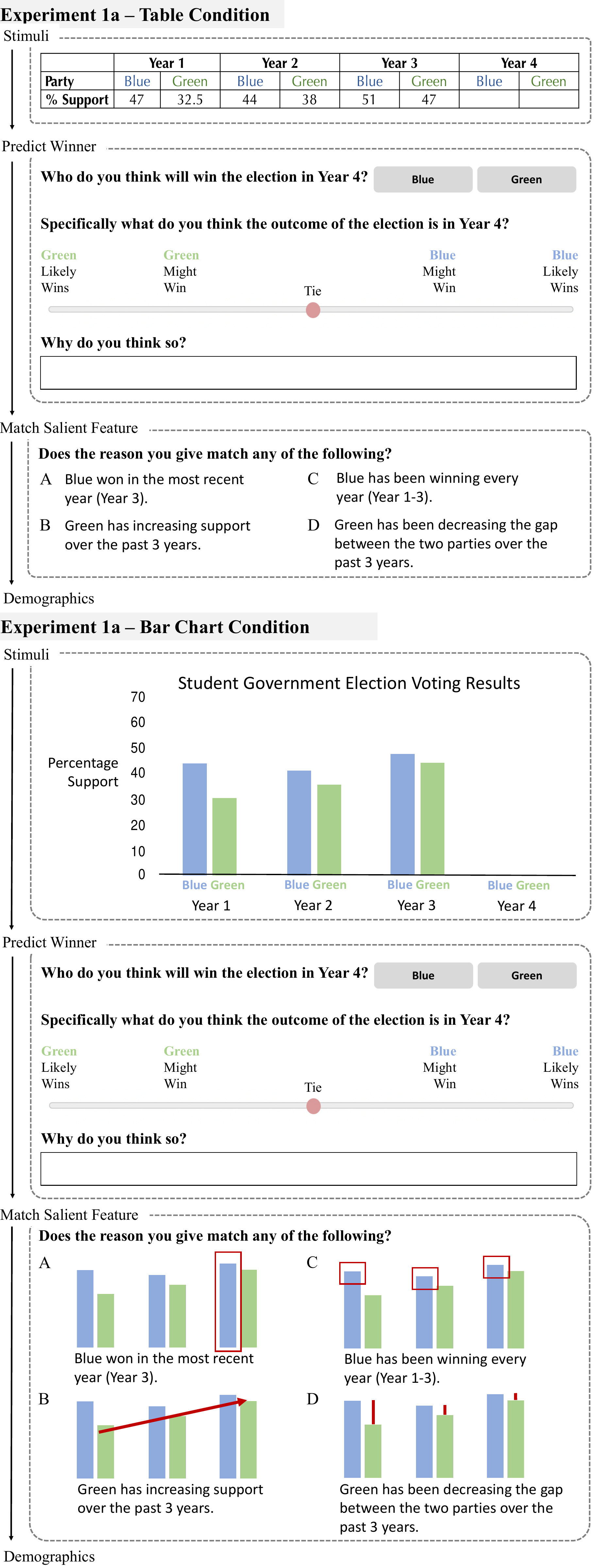}
\caption{Set-up for Experiment 1. Participants first read the table or the bar chart, predicted a winner and provided justification. Next, they matched their justification with one of four options we offered.}
\vspace{-5mm}
\label{fig:bar}
\end{figure}

Participants completed all tasks via surveys released through Qualtrics \cite{qualtrics2014qualtrics} on Amazon's Mechanical Turk.
They were randomly assigned to read the bar chart or table and then complete a decision task as shown in Figure \ref{fig:bar}.
The decision task asked them to predict which party would win the student government election in Year 4.
Participants reported their decisions in two ways: a binary forced choice (blue or green party) and a slider indicating the likelihood of their predicted victory.
Whether the two reporting methods yielded the same decision was used as one attention check.

The decision slider ranged from 0 to 50, where 0 indicated green likely wins, 25 indicated a tie, and 50 indicated blue likely wins, with the numerical values hidden from participants.
They also briefly explained their reasoning. 
On a separate page, participants matched their explanation to one of four choices (presented to the participants at the same time), shown in Figure \ref{fig:bar}.
Participants in the Table condition saw only the sentences, not the annotated charts.
These choices were collected from pilot studies in which participants indicated patterns that they saw in the bar chart.
If the participant could not match their reasoning to one of the choices, they were instructed to select ``other.''

\subsection{Stimuli}
The Bar Chart Condition in Figure \ref{fig:bar} shows the bar chart stimulus we used. It displays the results of a student government election between the blue and green parties over the past three years.
Y-axis shows percentage of support for the two parties.
Participants were asked, ``Which party will win in Year 4?''

The visualization contains competing data patterns that may be used to answer this question. 
We showed these patterns as choices under the `Match Salient Feature' section in Figure \ref{fig:bar}. 
One could notice that blue has won most recently (A) or every year from Year 1 to Year 3 (C), both suggesting that blue is more likely to win. 
One could alternatively notice that green has been gaining support (B) or that it has decreased the gap between itself and blue (D) from Year 1 to Year 3, concluding that green is more likely to win. 
The underlying data values from this bar chart were used to generate the table stimuli, as shown in the Table Condition of Figure \ref{fig:bar}, where the percentage support is 47, 44, and 51 for blue and 32.5, 38, and 47 for green from Year 1 to Year 3.
\change{We iteratively tested multiple values for each bar through pilot experiments to balance the saliency of the increasing pattern in the green bars and the differences between each pair of green and blue bars to obtain these final values. 
The exact values iterated can be found in the supplementary materials.}

\begin{figure}[h!]
\centering
\includegraphics[width=\linewidth]{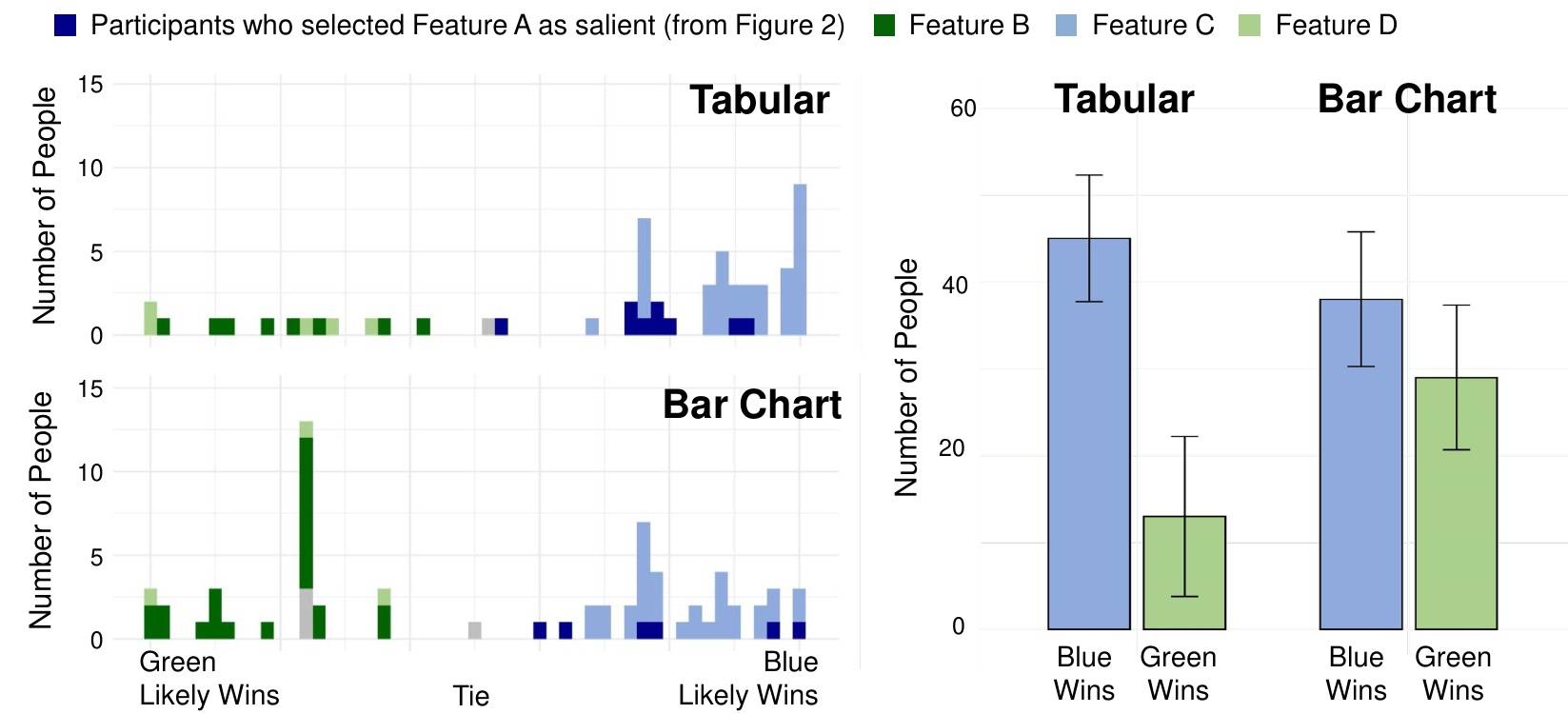}
\caption{Results from Experiment 1a. \textbf{Left}: distribution of slider responses on election outcome prediction for bar and table.
The color encodes the specific green or blue supporting features participants chose as their reasoning for their predictions. 
\textbf{Right}: prediction results for binary forced choice task from bar and table conditions. Error bars are calculated by estimating standard errors for proportions and then multiplied to a total sample size.}
\label{fig:exp1}
\end{figure}
\vspace{-2mm}

\subsection{Results}
Figure \ref{fig:exp1} (left) compares the conclusions that people drew from the bar chart versus the table reported using the slider (see Figure \ref{fig:bar}), where participants indicated the likelihood that either party would win in Year 4.
The slider response shows a right-skewed distribution for the table and a bimodal distribution for the bar chart.
We computed the Earth Mover's Distance (also known as 1-D Wasserstein distance \cite{schuhmacher2017package}) between the two distributions, which measures the similarity between them or the amount of `work' it takes to move points until the two are identical.
The larger the distance, the more different the distributions.
The Earth Mover's Distance between the two distributions is 8.79, which means we need to move each participant on average 8.79 units on the slider scale to make the tabular distribution identical to the bar chart distribution.
To get a better sense of the effect size, we compared this Earth Mover's Distance to the distance between the bar chart responses and a \textit{normal} distribution centered around Tie (x = 25) with a standard deviation that equals the pooled standard deviation of the bar chart and table response.
The distance turned out to be 4.39.
This suggests that the difference between the bar chart distribution and the tabular distribution is almost twice that between the bar chart distribution and a normal distribution.
 
Figure \ref{fig:exp1} (right) shows the results of the binary forced choice task. 
Participants were equally likely to report blue or green wins when reading the bar chart, but more likely to report that blue wins when reading the table ($\chi^2$ = 6.07, $Cramer's V$ = 0.22, $p$ = 0.014).
The color-coding in Figure \ref{fig:exp1} (left) represents the feature (see Figure \ref{fig:bar}, Matching Salient Feature section) participants picked to support their decisions.
Light blue maps to ``blue won in the most recent year (A).'' Dark blue maps to ``blue has been winning every year (C).'' Light green maps to ``green has increasing support over the past 3 years (B).'' Dark green maps to ``green has been decreasing the gap between the two parties over the past 3 years (D).'' 
Gray represents `others' - the participant indicated that none of the options captured their salient feature. 
Only 4.5\% of the responses fell in the `other' category and examples include ``I guessed" (details are in the supplementary materials).

From the color distribution in Figure \ref{fig:exp1} (left), we see that most people focused on blue-features when they predicted blue to win, and most people focused on green-features when they predicted green to win ($\chi^2$=120, $Cramer's V$=1, $p$<0.001), showing an association between the patterns that people notice and the conclusions that they draw. 

\subsection{Discussion}
Participants who were presented with the bar chart came to one of two conclusions, which they justified depending on whether they saw the blue supporting or green supporting feature as visually salient.
This result empirically demonstrates that visualizations are rhetoric devices that can elicit competing conclusions depending on what patterns readers focus on, \change{rejecting the null in favor of the alternative for the listed} \textbf{Hypothesis No.1} from Section \ref{Hypotheses}.
Furthermore, we found strong asymmetry in their conclusions between the table and bar chart, even within a trivially simple 2x3 dataset, \change{rejecting the null in favor of the alternative described in} \textbf{Hypothesis No.2}, which suggests visualization design can afford different interpretations.
The bar chart likely affords a clearer perception of the steady increase in the green bars, which can be seen as a single increasing slope, as opposed to requiring multiple arithmetic comparisons between blue and green (which leads to seeing blue as more popular throughout the years) within the table. 

The bimodal distribution of peoples' decisions with the bar chart echoes existing findings from economics, which suggests that people may possess some stable tendency to focus on specific types of patterns in data \cite{dominitz2011measuring}: a persistent type who believes that trends will persist, a reversion type who believes that trends will reverse, and a random-walk type who believes in no specific patterns.
Similar to existing work in the financial literature finding 'prototype' people profiles, we also observed three types of participants in this experiment: 
1) blue supporters who focused on the persistent dominance of blue parallel the persistent type, whom we will refer to as \textit{persistent position comparers}, 
2) green supporters who focused on the increasing trend of green (or the decreasing gaps between green and blue) parallel the reversion type, whom we will refer to as \textit{direction differentiators}, and 
3) others, such as people who believed in ties or saw salient features not corresponding with their predictions, whom we will refer to as \textit{random walkers}. 
We further investigate the stability of these tendencies in Experiment 1b, where we attempt to replicate current findings, and in Experiment 2, where we examine whether these three types of tendencies generalize across other types of visualizations. 

\section{Experiment 1b: Bar Chart Task Order Change} 

Experiment 1a asked participants to predict a winner before providing the data pattern they found salient. 
However, making the decision first might have led participants to pick a pattern that was consistent with their decision, \textbf{instead of having pattern salience drive the decision.} 
To strengthen the evidence for the direction of this relationship and to account for order effects, we reverse the order of these questions and tested only the bar chart condition, where responses showed far stronger bimodality, see Figure \ref{fig:exp1} for a comparison.

\subsection{Participants, Design, and Procedure}
We recruited 155 participants from Amazon's Mechanical Turk.
After filtering for workers following the same exclusion criteria \change{and compensation rate} from Experiment 1a, we are left with 141 individuals ($M_{age}$ = 33.40(9.91), 58 female participants).
Participants viewed the same bar chart as in Experiment 1a, but identified the most visually salient feature of the four choices (see the bottom panel in Figure \ref{fig:bar}) before completing the decision task on a new page, where they predicted the election winner in Year 4 using both the slider and the binary forced-choice response. 

\begin{figure}[h!]
\centering
\includegraphics[width = \linewidth]{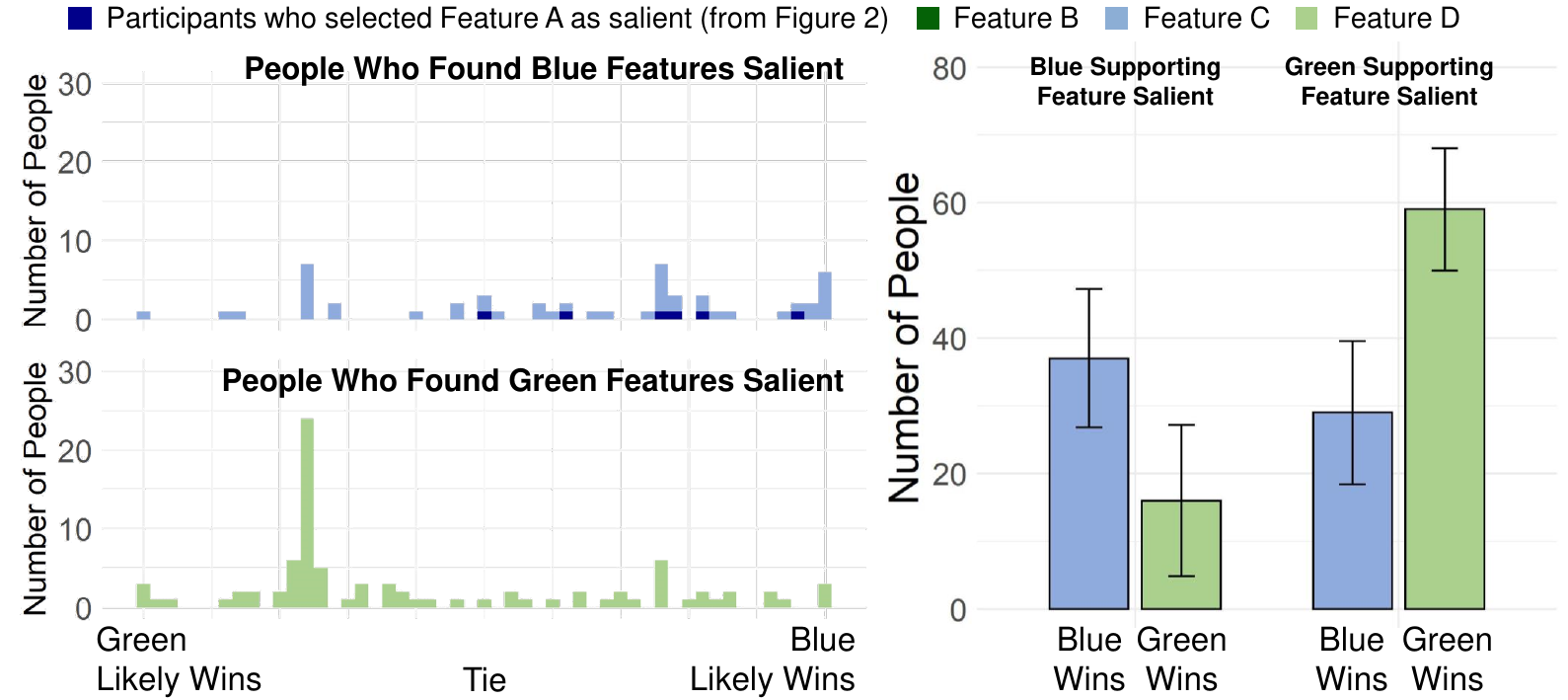}
\caption{\textbf{Left}: Distribution of slider response on election outcome prediction. The color encodes the specific green or blue supporting features they found salient, see Figure \ref{fig:exp1}. No participant indicated the pattern encoded by dark green as visually salient. \textbf{Right}: Summary of binary decisions, showing a congruence between salient patterns with predictions of which party would win. Error bars represent standard errors calculated by estimating standard errors for proportions and then multiplied to a total sample size.}
\label{fig:exp2}
\end{figure}

\subsection{Results}
We categorized participants based on the patterns they chose as visually salient, which can be either blue or green supporting features (see the Match Salient Feature section in Figure \ref{fig:bar}: options A and C are blue-supporting features, and options B and D are green-supporting features).
Overall, 37.6\% indicated blue supporting features as visually salient, and 62.4\% indicated green supporting features as visually salient.
Of the participants who indicated blue supporting features as visually salient, 69.8\% predicted blue to win, and 30.2\% predicted green to win on the binary forced choice response (as shown in the right panel of Figure \ref{fig:exp2}).
Of the participants who indicated green supporting features as visually salient, 67.0\% predicted green to win, and 33.0\% predicted blue to win.
Overall, participants who predicted blue to win were more likely to have indicated blue-supporting features as visually salient prior to making the prediction, and those who predicted green to win were more likely to have indicated green-supporting features as salient. 
A chi-square analysis shows a significant difference between identified salient features for people who made differing predictions ($\chi^2$ = 18.04, $V$ = 0.36, $p$ < 0.001). 

Figure \ref{fig:exp2} (left) shows the participant slider responses on election outcome likelihood, separated by features they found visually salient. 
On average, participants who found blue features salient were more likely to report blue to win (mean rating = $31.66$, SD = $14.16$), and participants who found green features salient were more likely to predict green to win (mean rating = $20.51$, SD = $13.68$).
The difference is statistically significant, with a large effect size (t = 4.58, p < 0.001, cohen's d = 0.80).
Note that the `tie' option translates to a numerical rating of 25.
Above 25 means they are blue-leaning, whereas below 25 means they are green-leaning.
The Earth Mover's Distance between the two distributions (blue salient vs. green salient) is 11.15, which means we need to move each participant on average 11.15 units on the slider scale to make the blue distribution identical to the green distribution.
This distance is larger than the distance between the bar chart distribution and table distribution of Experiment 1a, suggesting that the difference between predictions of people who find blue features salient and green features salient is even larger than that between people who see the bar chart and the table.



\subsection{Discussion}
This experiment replicated the bar chart results from Experiment 1a while ruling out potential ordering effects by first asking participants to indicate which patterns in the chart they find visually salient before making their prediction.
\change{This rejects the null hypothesis in favor of the alternative, which was described as \textbf{Hypothesis 1}.}
This also suggests there may be saliency-driven tendencies influencing what patterns people extract from data and the decisions they make from data.
But the current experiment is not causally manipulated, so we further test this effect by directly manipulating saliency in Experiment 3.

We also observed three types of behaviors in our responses: persistent position comparers who found blue features salient and predicted blue to win, direction differentiators who found green features salient and predicted green to win, and random walkers who were not strong supporters of either blue or green. 
We designed a subsequent experiment to examine whether these three types of behaviors persist and whether viewers would consistently extract similar types of features even in other visualization types.
\section{Experiment 2: Bar and Line Charts} 
Experiments 1a and 1b showed that in bar charts, people can perceive varying patterns to be salient. 
There seems to be evidence suggesting people can fall into one of three categories in terms of their tendency to find specific patterns as salient: persistent position comparers find the dominant lead of the blue bars visually salient, direction differentiators find the increasing trends salient, and random walkers see both equally.
Furthermore, their perceived saliency of data patterns can drive the decisions they make with data.
We test whether this effect generalizes to other visualizations and whether the same individual would find similar types of patterns salient across multiple visualizations using a line chart as a test case in addition to the bar chart.

\begin{figure}[h]
\centering
\includegraphics[width = 0.88\columnwidth]{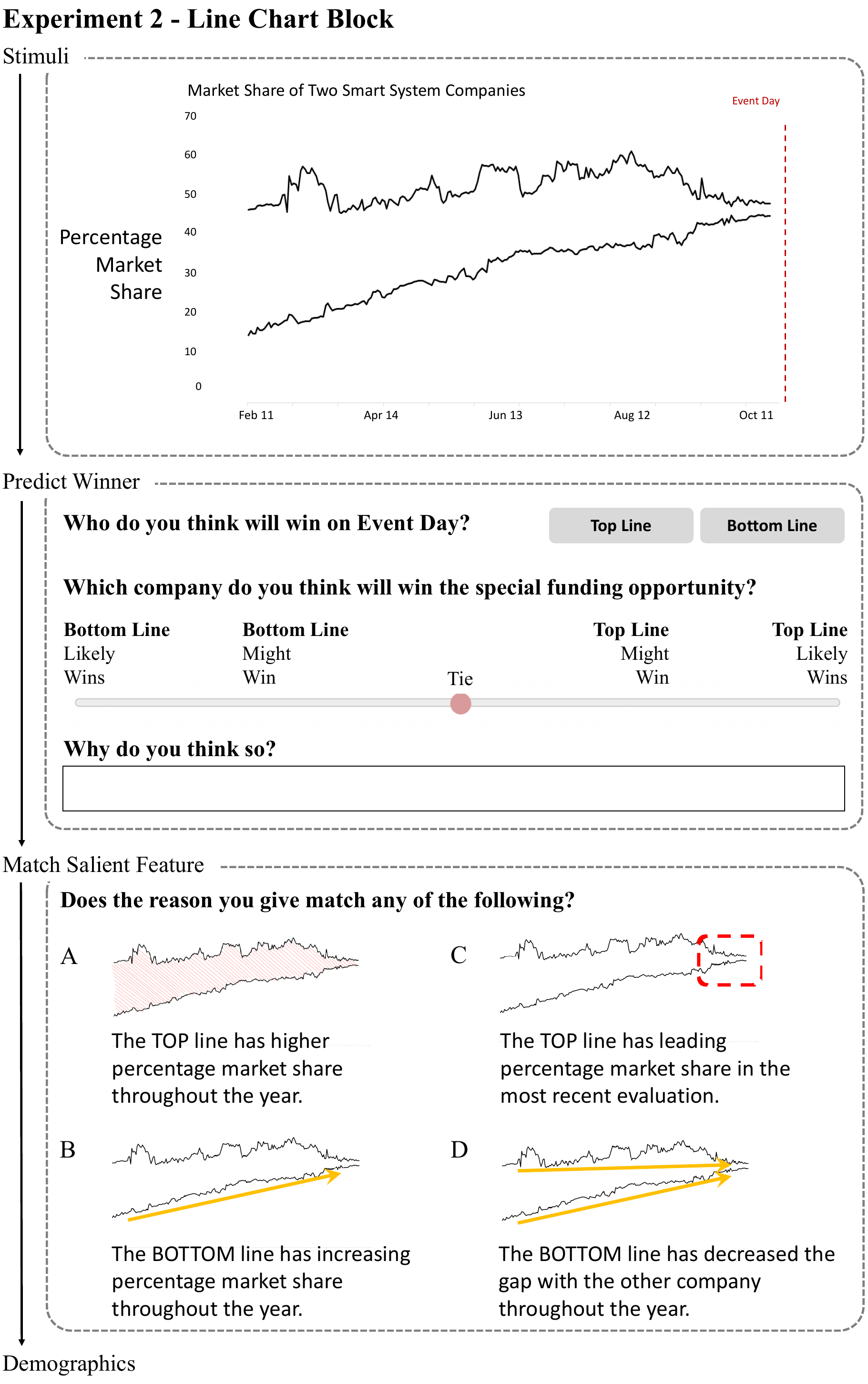}
\caption{Procedure for the line chart portion of Experiment 2. 
}
\label{fig:line}
\end{figure}
\vspace{-5mm}

\subsection{Stimuli}
The top panel in Figure \ref{fig:line} shows the line chart used in Experiment 2.
It is designed to resemble the bar chart from Experiments 1a and 1b, depicting a competition between two companies (top line and bottom line) over time (x-axis) for a higher percentage of market share (y-axis). 
The top line parallels the blue party in the bar chart.
As shown in the `Match Salient Features' panel in Figure \ref{fig:line}, one might notice that the top line has a steadily high market share throughout the year (A), or that it is leading most recently (C).
The bottom line parallels the green party in the bar chart.
One might notice that the bottom line slopes upwards (B), or that it has drastically decreased the gap with the top line (D).
The participant is prompted to decide which line will win by having a higher market share by Event Day as the parallel to deciding which party will win the election in Year 4 from the bar chart.

\subsection{Participants, Design, and Procedure}
We recruited 294 participants from Amazon's Mechanical Turk.
After filtering for workers following the same exclusion criteria from Experiment 1a and 1b, we are left with 271 individuals ($M_{age}$ = 36.78(11.56), 47 female).
In this experiment, participants read both the bar chart and line chart in a random order following procedures from Experiment 1 for the bar chart and Figure \ref{fig:line} for the line chart, separated by a distractor task on word clouds. 
Participants predicted the winner, provided their reasoning, and identified salient features, as they did in Experiment 1. 
At the end of the experiment, participants reported whether they thought the two visualizations were related in any way and speculated about the purpose of the experiment. 

\begin{figure*}[h!]
\centering
\includegraphics[width = \linewidth]{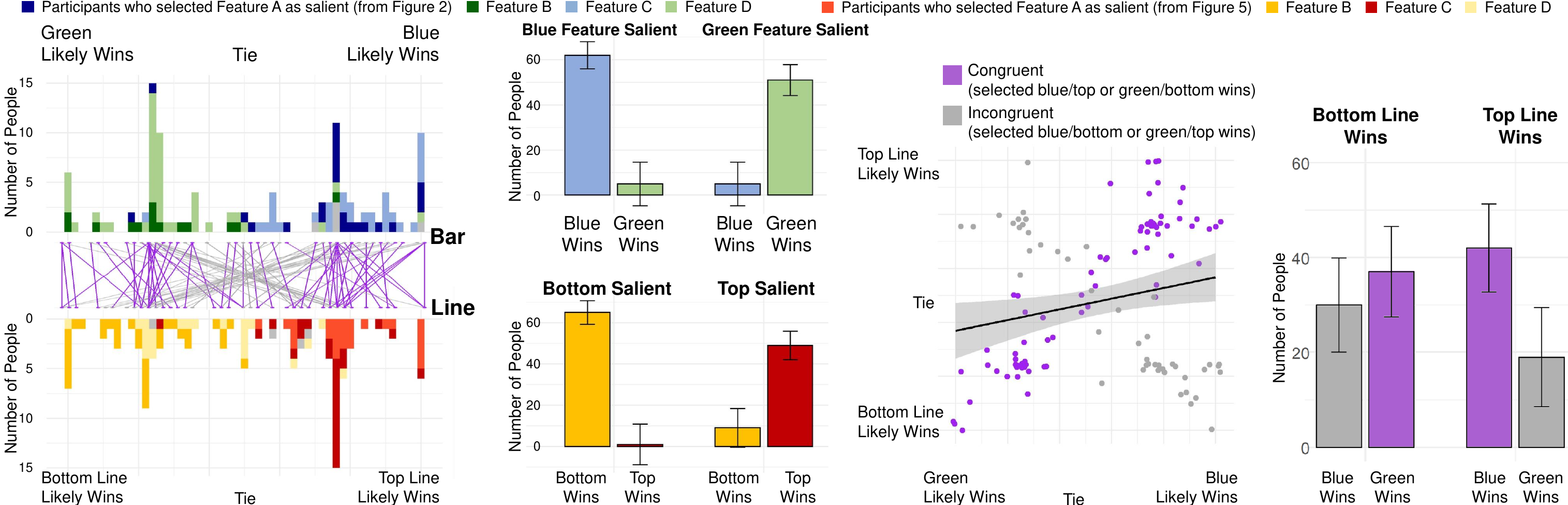}
\caption{\textbf{Left}: Histograms showing the distribution of bar chart election outcome prediction and line chart market share prediction.
The color encodes the specific pattern participants indicated as salient, following the same color schemes as in Experiment 1.
The lines connecting the two histograms show what the same participants predicted across the two charts. Purple lines represent those who made congruent predictions (blue/top or green/bottom), and gray lines represent those that made incongruent predictions (blue/bottom or green/top).
\textbf{Middle Left}: Binary response results on the predicted winner by chart features they found salient.
\textbf{Middle Right}: Scatterplot showing how bar chart prediction slider values correlate with those for the line chart. Overall, more participants made congruent predictions, with R = $0.21$.
\textbf{Right}: Binary response results for the bar chart and the line chart. Error bars represent standard error calculated by estimating standard errors for proportions and then multiplied to a total sample size.
}
\label{fig:noisyLine_Bar}
\end{figure*}

\subsection{Results: Overall Decision Distribution}
Slider responses for the \textbf{bar chart} replicated the findings from Experiments 1a and 1b, in which participants were evenly split in predicting whether blue or green would win, as shown in Figure \ref{fig:noisyLine_Bar}. 
The binary forced choice responses also replicated the participants' tendency to identify blue supporting features as visually salient when they predicted blue party win and identify green supporting features as salient when they predicted green to win ($\chi^2$ = 85.98, $V$ = 0.84, $p$ < 0.001).

For the \textbf{line chart} (shown in the histogram in Figure \ref{fig:noisyLine_Bar}), participants were also evenly split in predicting whether the top or bottom line would win.
This shows that the ambiguous nature of visualizations to afford opposing interpretations can be generalized to line charts.
Mapping of yellow and red colors onto the histogram is analogous to the mapping for the bar chart. 
Dark yellow maps to ``bottom line increased percentage market share throughout the year (B).'' 
Light yellow maps to ``bottom line decreased the gap between the two lines throughout the year (D).''
Dark red maps to ``top line is leading in the most recent evaluation (A).'' 
Light red maps to ``top line has higher percentage market share throughout the year (C).''
The binary forced-choice responses further support the generalizability that the patterns in data participants find visually salient can drive the decisions they make.
Most participants focused on bottom line supporting features when they predicted the bottom line to win and top line supporting features when they predicted the top to win ($\chi^2$ = 88.31, $V$ = 0.84, $p$ < 0.001).

We calculated the Earth Mover's distance between the bar and line chart slider distributions to be 3.52.
This means we need to move each participant on average 3.52 units on the slider scale to make the bar and line distributions identical.
This suggests that the difference between people's overall reactions to the line and the bar chart is smaller than the difference between the bar chart and the table from Exp 1a. 

\subsection{Results: Prediction and Perceived Patterns}
The line and bar charts were designed to be analogous, such that the top line resembles the blue party and the bottom line resembles the green party.
We will refer to a participant as \textit{decision-congruent} if they made similar predictions across the two charts.
That is, they are decision-congruent if they predicted blue/top or green/bottom to win, and \textit{decision-incongruent} otherwise.

Critically, more participants were congruent ($\chi^2$=7.52, $V$=0.24, $p$<0.01), as shown in the right-most panel in Figure \ref{fig:noisyLine_Bar}.
With the binary response, those who predicted green to win were also more likely to predict the bottom line to win, and those who predicted blue to win were also more likely to predict the top line to win.
The scatterplot in Figure \ref{fig:noisyLine_Bar} corroborates this finding with the slider scale responses. 
There is a small effect for decision congruence, as shown with the positive slope of the regression line in the scatterplot.

\subsection{Discussion}
This experiment shows that people can see different patterns as visually salient in line charts and make competing decisions, just as they would with bar charts, suggesting that the effects we observed in Experiments 1a and 1b generalize to other visualization types, again supporting \textbf{Hypothesis 1}.
We can also compare the prediction distributions of bar charts and line charts in this experiment to that of bar charts and table in Experiment 1 to infer that tables are less effective at making upward trends visually salient.

We again observed the three types of people in this experiment in terms of what data patterns they tend to focus on.
Persistent position comparers saw blue-supporting features and top line-supporting features as salient.
Direction differentiators saw green-supporting features and bottom line-supporting features as salient.
Both types behave consistently in what they find visually salient across visualizations and tend to make congruent decisions, providing support for \textbf{Hypothesis 3}, \change{rejecting the null hypothesis that there does not exist intra-personal consistencies}.
Random walkers are less consistent, and as shown in the gray data points in Figure \ref{fig:noisyLine_Bar}, they tend to be incongruent in the patterns they indicated as visually salient and were also decision-incongruent.

However, previously identified salient features (e.g., a growing trend) may have priming effects on how participants see subsequent visualizations and make decisions, resulting in more congruency.
We provide further speculations on this effect and discuss future opportunities to further tease apart the mechanisms in Section \ref{future}.

We also designed a follow-up experiment that leverages highlights and annotations to manipulate the perceived saliency of certain patterns.
The experiment will allow us to test the extent to which visual saliency and personal preferences in data pattern extraction play a role in what people see and decide.



\begin{figure*}[h]
\centering
\includegraphics[width = 0.91\linewidth]{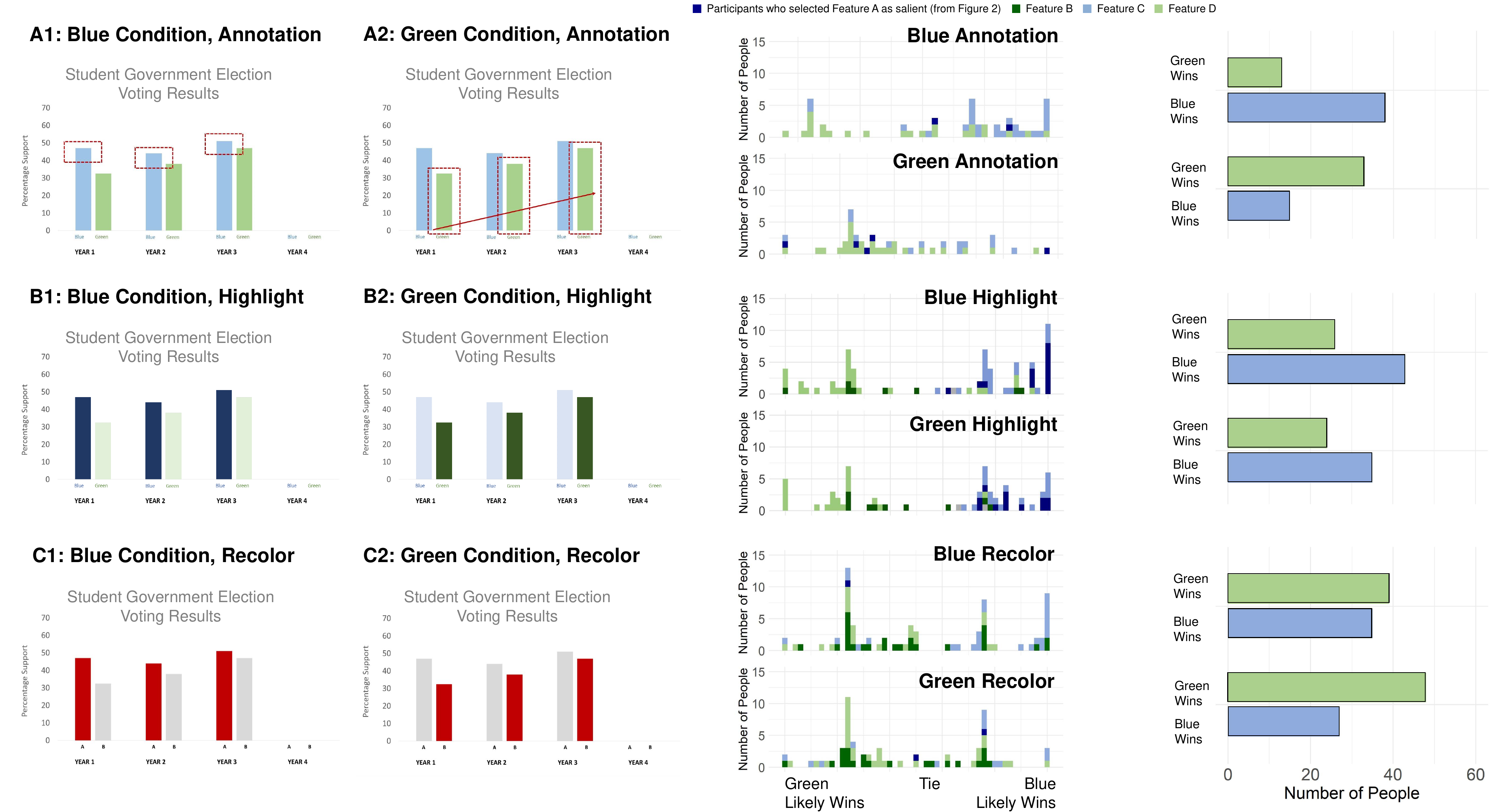}
\caption{\textbf{Left}: Stimuli shown to participants: annotation, highlight, or recolor of blue supporting or green supporting features. 
\textbf{Middle}: Distribution of participant responses for the annotation, highlight, and recolor conditions. The color encodes which specific patterns participants indicated as salient, following the same color schemes as in Experiment 1. 
\textbf{Right}: Prediction results for the annotation, highlight, and recolor conditions.
 }
\label{fig:Experiment3New_Results_Prime_Annotation_Highlight}
\end{figure*}

\section{Experiment 3: Annotations and Highlighting} 
Experiments 1a, 1b, and 2 showed the potential effects of perceived visual saliency and personal preferences (e.g., persistent position comparers) on what people see in a visualization and their eventual decisions. 
To seek evidence of a stronger causal role of the two, we created this experiment to test the effect of two manipulations commonly used to emphasize data patterns: annotation and highlighting. 
We hypothesize that if visual saliency plays a significant role in deciding what patterns people see and what they decide, we would observe people who see the annotated/highlighted patterns as salient and make congruent decisions to the manipulation (e.g., predicting blue to win when the annotation highlights blue's winning streak). 
However, if personal preferences play a role (e.g., persistent position comparers will always be persistent position comparers), we would observe that annotations/highlighting techniques have a negligible effect.


\subsection{Participants, Stimuli, Design, and Procedure}
We recruited 455 participants for Experiment 3.
After applying the same exclusion criteria as previous experiments, we ended up with 376 participants ($M_{age}$ = 32.46(9.79), 179 female participants).

As shown in Figure \ref{fig:Experiment3New_Results_Prime_Annotation_Highlight}, we created a 2x3 design with two storytelling techniques that emphasize patterns in data: annotation (A1, A2) and highlighting (B1, B2), including recolored highlighting (C1, C2).
For each technique, we created two conditions: one emphasizing a blue-supporting pattern, and one emphasizing a green-supporting pattern.
For the annotation/blue condition, we annotated blue's winning streaks.
For the annotation/green condition, we used a trend-line to draw attention to green's upward trend.
For the highlight/blue condition, we decreased the visibility of the green bars.
For the highlight/green conditions, we decreased the visibility of the blue bars.
For the recolor/blue condition, we colored the blue bars to a salient red color and the green bars to a less salient gray color.
For the recolor/green conditions, we colored the blue bars a less salient gray and the green bars a more salient red. 
\change{This recoloring condition allows us to test the generalizability of the effect of a color change. We chose bright red in contrast to light gray to maximize the visual saliency of the highlighted party.}
\change{To check the validity of our saliency manipulation, we ran our stimuli through the visualization saliency model proposed by Matzen et al. \cite{matzen2017data} and found the saliency map in line with our design intents. 
Highlighted and annotated areas are predicted by the saliency model to be more visually salient to a viewer. 
The saliency maps for these stimuli can be found in the supplementary materials.}

Participants were randomly assigned to read one of the bar charts shown in Figure \ref{fig:Experiment3New_Results_Prime_Annotation_Highlight}. 
They then indicate the party they predict to win via binary forced choice, and on a separate page, the features they find visually salient, followed by the slider prediction task of likelihood, following a similar procedure as that in Experiment 1.

\subsection{Results: 3A Annotation}
Participants in the annotation/blue condition more often indicated blue to win, and vice-versa for those in the annotation/green condition ($\chi^2$ = 17.63, $V$ = 0.43, $p$ < 0.001), see Figure \ref{fig:Experiment3New_Results_Prime_Annotation_Highlight} (right).
We again compared the distribution of slider scale responses via Earth Mover's Distance and found that in order to make the two distributions identical, we would need to move each participant on average 13.72 units.
Compared to Earth Mover's Distances from other experiments, the two distributions in this experiment are the most different, as shown in Table \ref{EMD}.
This result suggests that emphasizing visual saliency of data patterns via annotations can influence viewer data perception and decision.

\begin{table}[h]
\begin{tabular}{l|l|l|l|l|l|l}
\hline
\textbf{Experiment}             & \textbf{1a} &\textbf{ 1b} & \textbf{2} & \textbf{3A}  & \textbf{3B} & \textbf{3C} \\ \hline
\textbf{EMD} & 8.79    & 11.15       & 3.52     & 13.72    & 1.88    & 4.91  \\ \hline
\end{tabular}
\caption{The Earth Mover's Distance from all experiments.}
\vspace{-5mm}
\label{EMD}
\end{table}

\subsection{Results: 3B Highlight}
Participants in the blue condition did \textit{not} more often indicate blue to win, and those in the green condition did not more often indicate green to win ($\chi^2$ = 0.15, $V$ = 0.034, $p$ = 0.69), as shown in Figure \ref{fig:Experiment3New_Results_Prime_Annotation_Highlight} (right).
We again compared the distribution of the slider scale responses via Earth Mover's Distance and found that in order to make the two distributions identical, we would need to move each participant on average 1.88 units.
Compared to the Earth Mover's Distances from other experiments, the two distributions in this experiment are the \textit{least} different, as shown in Table \ref{EMD}.
This result failed to provide evidence that emphasizing the visual saliency of data patterns via highlighting influences viewer data perception and decision-making. 
This suggests that the salience manipulation of highlighting the bars darker blue/green was not effective in making the corresponding blue/green patterns stand out, and participants might have relied on their personal preferences when making their decisions. 

\begin{figure}[h]
\centering
\includegraphics[width = \columnwidth]{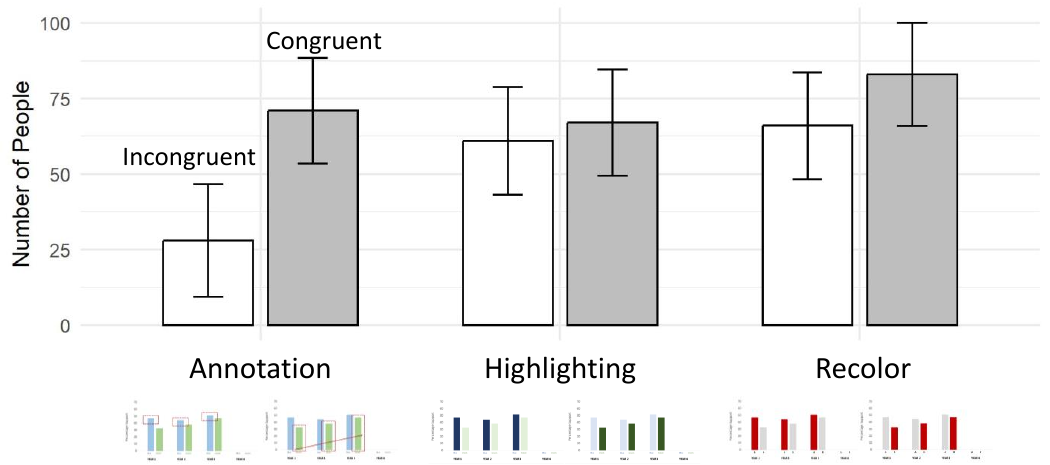}
\caption{Number of people who made congruent and incongruent predictions with the annotated and highlighted visualizations.
}
\label{fig:Exp3_congruentIncongruent_perManipulation}
\end{figure}

\subsection{Results: 3C Highlight with Recoloring}

Participants in the blue condition did \textit{not} more often indicate blue to win, and those in the green condition did not more often indicate green to win ($\chi^2$ = 1.96, $V$ = 0.11, $p$ = 0.16), as shown in Figure \ref{fig:Experiment3New_Results_Prime_Annotation_Highlight} (right).
We again compared the distribution of the slider scale responses via Earth Mover's Distance and found that in order to make the two distributions identical, we would need to move each participant on average 4.91 units.
This result again failed to provide evidence that emphasizing the visual saliency of data patterns via highlighting a different color pallet influences viewer data perception and decision-making, corroborating our observations from the highlight condition (3B).

\subsection{Comparing Annotation, Highlighting, and Recoloring}
We examine the strength of influence of annotation, highlighting, and recoloring by comparing the number of participants who provided a decision that is congruent with the prime.
For example, a response is considered congruent if the annotation emphasizes that blue has been winning every year (blue condition) and the participant predicted blue to win.
Figure \ref{fig:Exp3_congruentIncongruent_perManipulation} shows the number of congruent and incongruent responses.
Overall, annotation is more impactful than highlighting and recoloring.
Comparing the results from 3A, 3B, and 3C, participants were more likely to draw conclusions congruent to the annotated visual primes in the annotation condition (3A), 
but were equally likely to draw incongruent and congruent conclusions in the highlighting (3B) and recoloring conditions (3C).
The results suggest that storytelling techniques aimed at increasing saliency are not equally effective.
This observation \change{rejects the null in favor of} both \textbf{Hypothesis No.4} and \textbf{Hypothesis No.5}, suggesting that visual saliency may play a role, and together with the results from Experiments 1 and 2, that personal preferences likely also play a role when saliency measures are not strong, corroborating existing work showing that individual differences such as locus of control \cite{ziemkiewicz2012visualization} and personality traits \cite{ottley2015personality}. 
However, explanations other than personal preference may also be plausible. 
We discuss several possibilities in Section \ref{future}. 



\subsection{Discussion}
The results suggest that both visual saliency and personal preferences play a role in visual pattern extraction and data decision-making.
We see evidence that annotation is a stronger cue than highlighting/recoloring to influence what people see and decide from data, \change{rejecting the null in favor of} \textbf{Hypothesis No.5}.
When the visual saliency of patterns is significantly increased by stronger cues like annotation, viewers tend to draw congruent decisions.
When the cue is less effective (in this case, the highlighting and recoloring manipulation), viewers lean towards their personal preferences to extract patterns and make decisions.

Existing literature and best practices in data storytelling encourage designers to leverage annotation and highlighting to focus visualization readers on the `right' story \cite{ajani2021declutter}, but this experiment demonstrates that they are not equally effective.
This motivates us to reflect on our data storytelling practices and the power we hold as visualization designers to impact what people see from data and what decisions they make from data.
See further discussions in Section \ref{future}.


\section{Summary}

\noindent \textbf{Experiment 1a} showed that people find different patterns salient and make different decisions based on visualization type, \change{rejecting the null in favor of} \textbf{Hypothesis No.2}. For the bar chart, 
people saw different patterns from the same visualization, \change{rejecting the null in favor of} \textbf{Hypothesis No.1}.

\noindent \textbf{Experiment 1b} replicates the bar chart study from Experiment 1a while reversing the task order to rule out potential ordering and priming effects. It further demonstrates that visualizations are not agnostic: the same visualization can elicit different decisions, \change{rejecting the null in favor of} \textbf{Hypothesis No.1}.
This suggests that visual saliency can influence the patterns people see and decisions they make.

\noindent \textbf{Experiment 2} replicates the bar chart stimulus from Experiment 1 and additionally shows participants a line chart, which was designed with visual features conceptually similar to the bar chart.
Across both charts, participants tended to be intra-personally consistent and saw conceptually similar patterns as visually salient.
They also made similar decisions in both charts, \change{rejecting the null in favor of} \textbf{Hypothesis No.3}, which suggests that personal preferences also play a critical role in determining what people see and decide.

\noindent \textbf{Experiment 3} emphasized data through annotation and highlighting/recoloring. Annotation had a stronger impact than highlighting/recoloring on the patterns people see and their decisions, \change{rejecting the null partially in favor of} \textbf{Hypothesis No.5}. 
This also provides evidence that both visual saliency of data patterns and personal preferences take part in influencing what people see and decide, \change{rejecting the null in favor of} \textbf{Hypothesis No.4}.

\section{Discussion}
We demonstrate that visualizations 
can be interpreted similarly to ambiguous figures \cite{attneave1971multistability} 
such that two people looking at the same dataset and focus on different patterns to come to opposing conclusions.
There exists individual differences in what patterns people tend to focus on and what decisions they make, corroborating findings from existing work in behavioral economics, psychology, and visualizations (e.g., \cite{dominitz2011measuring}, \cite{torsney2015visualization}, \cite{ziemkiewicz2012visualization}, \cite{ottley2015personality}, \cite{gaba2023my}, \cite{franken2005individual}). 
Our participants fall into three types: \textit{persistent position comparers}, \textit{direction differentiators}, and \textit{random walkers}. 
Persistent position comparers compare the overall positions of all entities in the visualization, finding overarching dominance to be salient. 
These people tend to support the historical winner (the blue party in the bar chart and the top line in the line chart) and predict their winning streak to continue.
\textit{Direction differentiators} find increasing trends to be salient and tend to support the underdogs gaining momentum to reverse. 
These people tend to believe the increasing trend in the green party's popularity and bottom line`s market share will reverse the outcome. 
\textit{Random walkers} did not adhere to either principle when extracting patterns across multiple charts.
Most of our participants seem to be either a position comparer or a direction differentiator.
They sought out similar patterns (e.g., growing trends) and formed congruent conclusions (e.g., the underdog will win) across both charts, which implies some level of intra-personal consistency. 
This is consistent with findings in social psychology suggesting that people tend to be intra-personally consistent \cite{funder1991explorations}, likely to resolve internal cognitive dissonance \cite{harmon2019introduction} and reduce cognitive effort during tasks \cite{kool2010decision}. 

\begin{figure}[h]
\centering
\includegraphics[width = \linewidth]{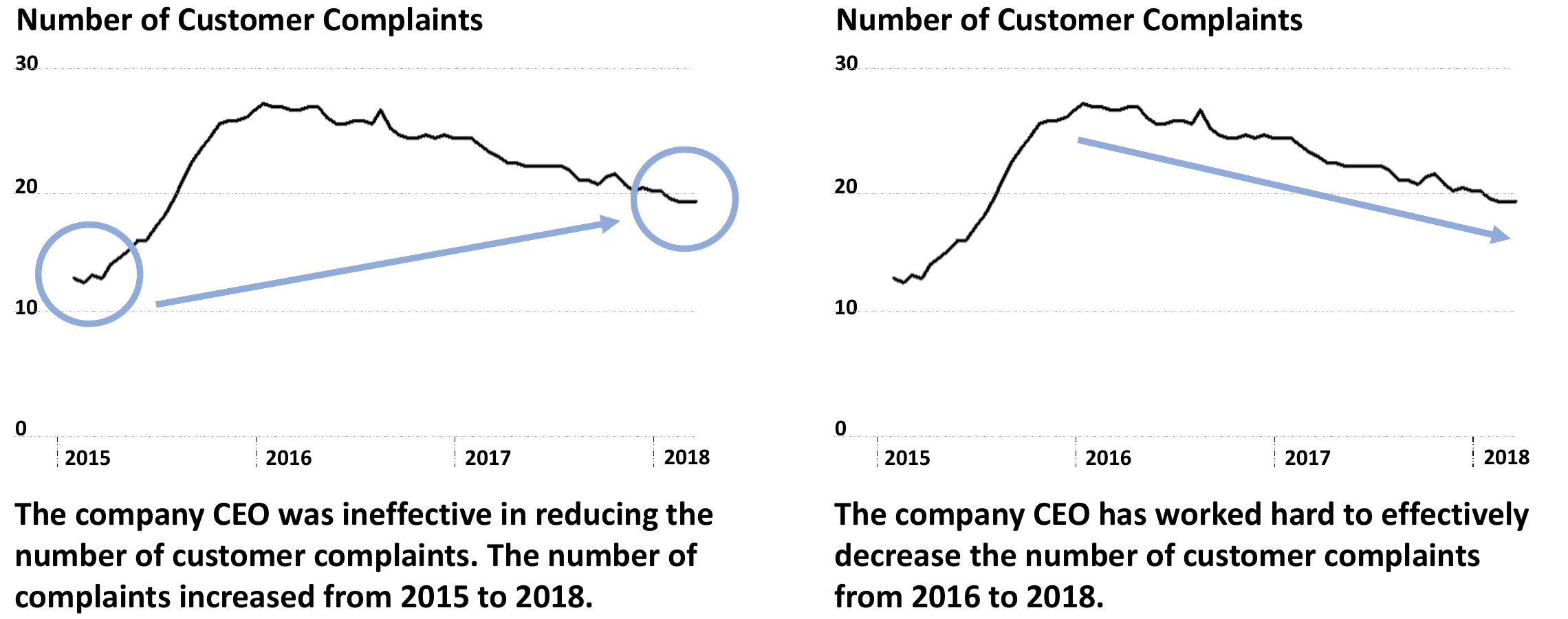}
\caption{Adaptation of a line chart from journalists (NYTimes Upshot) \cite{NYT_Example} \change{that demonstrates two annotations highlighting two perspectives. The line chart depicts the number of customer complaints overtime under the current leadership and can be seen either as increasing or decreasing.}}
\label{fig:nytExample}
\end{figure}

Experiment 3 demonstrated that when visualizations are annotated to emphasize certain patterns, people tend to see them as more salient and make corresponding conclusions.
We observed a weaker effect with highlighting and recoloring. 
These results suggest that 
design choices can play a critical role in determining what people see and decide beyond what people naturally find salient, and the power of annotation is especially strong.
Annotation is sometimes used by journalists to tell stories with data \cite{NYT_Example}. 
Figure \ref{fig:nytExample} \change{is an example of an ambiguous figure adapted from New York Times Upshot \cite{NYT_Example}, such that, depending on whether the reader takes a pro-leadership or an anti-leadership perspective, they could focus on different patterns in the visualization to draw different conclusions.} 
Oppositions of the leadership might notice an increase in overall customer complaints from 2015 to 2018 while supporters of the leadership might see a decrease in complaints from 2016 to 2018. 

\section{Design Implications and Guidelines}
Two people looking at the same visualization can draw different conclusions.
To focus people on the intended message in a visualization, designers leverage the power of data storytelling techniques such as annotation and highlighting.
However, there exists little existing empirical evidence demonstrating the effectiveness of these techniques on swaying reader decisions, despite widespread support from practitioners \cite{ajani2021declutter}.
The present paper serves as an empirical demonstration that storytelling techniques can be very powerful, especially direct annotation on a visualization (more so than highlighting).
We can build recommendation systems that help people more effectively explore, analyze, and present data by predicting people's takeaways from a given design \cite{gaba2022comparison, xiong2021visual}.
Because visual saliency and individual differences both play a role in predicting what people see and decide from data, we can build models incorporating these two factors to improve visualization recommendation systems.
Researchers already began modeling human visual attention when interacting with data visualizations \cite{fosco2020predicting,bylinskii2016should, bylinskii2018different}.
These models can not only be used to inform visualization design but also contribute to an intelligent visual analytic tool.
Additionally, these models can be used to offer explanations as to why people extract certain information from data, revealing underlying perceptual and cognitive mechanisms in visual data interpretation \cite{li2022predictable, bose2020decision}.
Furthermore, visualization researchers can leverage these models to identify biases in people's viewing patterns to inform the design of bias-mitigating tools \cite{xiong2022seeing, mantri2022how}.

Another point to consider is the ethical implications of leveraging techniques like annotation; it may sometimes be appropriate to avoid such manipulations. 
For example, if viewers desire the freedom to explore data without getting biased, or if they want to develop new ideas from the visualization, showing people annotations would taint their perspectives and make their interpretations less impartial.
We should reflect upon the implications of data storytelling techniques that leverage annotation and highlights and consider generating ethical data storytelling guidelines to empower visualization readers to critically think about data stories they read, rather than becoming tunnel visioned by focusing on only one side of the story.

\section{Limitations and Future Directions}
\label{future}
In the present study, participants performed two tasks: indicating the feature they found salient and predicting who they thought would win. 
While the order of the two tasks has been counterbalanced in Experiments 1a and 1b to show consistent results, we do not yet know whether there exists a causal relationship between the two, and in which direction the causal arrow points. 

The annotation condition from Experiment 3 suggests that there potentially exists a causal relationship between what people find salient and the decisions they make, but this causal relationship can be mediated by personal preferences when the saliency cue is not strong enough, as observed in the highlighting condition.
Future studies should further investigate the extent to which this causal relationship is perceptually driven (bottom-up), or individual difference driven (e.g., they are persistent position comparers).
We offer some discussions to hopefully inspire extensions of this research direction.

Viewers may, through a bottom-up process, find certain features and patterns visually salient, and then form a belief to drive their decisions.
For example, a participant looking at the bar chart on the student government elections may notice that the blue bars are overall taller and more noticeable.
The participant's attention is drawn to these tall blue bars. They did not even notice that the green bars are increasing over time. 
Their impression of the large blue bars subsequently led them to form a belief that the blue bars are so tall that blue will likely win. Experiment 3 shows that, to a degree, annotations could make some patterns more perceptually salient and lead to differing decisions.
On the other hand, viewers could exert their top-down attentional control and pick patterns that support their pre-existing beliefs. 
For example, a participant who believes that the general preferences of voters do not change over time may look for patterns in the data that support this belief, notice that the blue party has been historically victorious, and report the pattern of blue's dominance as salient. 
This may also be an instance of confirmation bias, as the viewer has already made up their mind and is merely looking for evidence to support their belief. 
We see evidence supporting this belief-driven process from the curse of expertise in visual data communication---once a data viewer sees a salient feature in a visualization because of their background knowledge, they tunnel-vision on the pattern, assuming everyone else would also see that feature as most visually salient while ignoring other potential patterns in the visualization \cite{xiong2019curse}. 
Both processes could have played a role in our experiment. 
This may imply that bottom-up processes can interact with top-down processes in data interpretation to form a feedback loop, creating a circular causal chain similar to the ``chicken or the egg?'' question. 
Future work could further tease these two processes apart and asses their effect in real-world visual data communication instances such as the one in Figure \ref{fig:nytExample}.


Furthermore, although the line chart in Experiment 2 was designed to parallel the bar chart, there was a drop in the top line towards the end that was not present in the blue bars. 
This difference could have led to differences in viewer perception and prediction.
While we were able to find consistencies in viewer behaviors nonetheless, the result might be on the conservative side.
We encourage future work to test charts that are more and less similar to each other across other types of visual representations to evaluate the generalizability of our findings.
\change{Moreover, future work could test more combinations of colors for groups of bars (including hue, contrast, and saturation differences), as well as other forms of annotation and highlighting methods, for even more generalizability across design choices.}

As we increasingly rely on data to understand, communicate, and make decisions, we need to further understand how our brains work to extract critical values, statistics, and patterns needed to make decisions about data.
This will empower us to design visualizations that increase data comprehension and effective communication. 
As Klein and O'Brien pointed out, when making decisions, people use less information than they think \cite{klein2018people}. 
It is critical to consider the implications of visualization design to facilitate data-driven, instead of motivation-driven or saliency-driven decisions with data. 



\acknowledgments{%
The authors would like to thank Chase Stokes and Akira Wada for their feedback. This work is partly funded by NSF awards IIS-2237585, IIS-1901485, and IIS-2311575.%
}

\balance
\bibliographystyle{abbrv}

\bibliography{reference}
\end{document}